\documentclass[aps,prl,showpacs,amsmath,amssymb,
	preprint,onecolumn]{revtex4}

\usepackage[dvips]{graphicx}
\usepackage{dcolumn}
\usepackage{bm}
\usepackage{epsf}

\begin{document}

\title{Synchronized dynamics of cortical neurons with time-delay feedback}

\author{Alexandra S. Landsman$^1$, Ira B. Schwartz$^1$}
\affiliation{$^1$US Naval Research Laboratory, Code 6792, Nonlinear Systems Dynamics Section, Plasma Physics Division, Washington, DC 20375}
\email{<alandsma@cantor.nrl.navy.mil>}

\begin{abstract}
The dynamics of three mutually coupled cortical neurons with time delays
in the coupling are explored numerically and analytically. The neurons are
coupled in a line, with the middle neuron sending a 
somewhat stronger projection
to the outer neurons than the feedback it receives, to model for instance the
relay of a signal from primary to higher cortical areas.
For a given coupling architecture, the 
delays introduce correlations in the  time series at the time-scale of
the delay.  
It was found that the middle neuron leads the outer ones by the delay
time, while the outer neurons are synchronized with zero lag times.
Synchronization is found to be highly dependent on the
synaptic time constant, with faster synapses increasing both
the degree of synchronization and the firing rate.  Analysis
shows that pre-synaptic input during the inter-spike interval 
stabilizes the synchronous state, even for arbitrarily weak coupling, 
and independent of the initial phase.
The finding may be of
significance to synchronization of large groups of cells in the cortex
that are spatially distanced from each other.
\end{abstract}

\maketitle

\section{\label{sec:intro} I. Introduction}

There is a significant
amount of research showing that spike coincidence of neurons encodes
information.
Synchronization of neural activity has been involved in such important
phenomena as ability to recognize objects (by binding different attributes) \cite{gerstner},
olfactory discrimination \cite{stopfer}, and has even been proposed as
one of the neural correlates of consciousness \cite{crick95}.
Abnormal neural synchronization has been linked to the onset of epileptic
attacks.  

Evidence of long range correlation leading to synchronized clusters
has been observed in the visual cortex \cite{EngelKKS91} where
synchronized oscillatory responses are conjectured to contribute to the
binding of distributed features in a visual scene. Further research in
\cite{KonigES95} on the cat visual cortex supports that
oscillatory activity and firing patterns can contribute to long range
synchrony among cortical cluster of neurons. Other mechanisms
involving gamma oscillations are also conjectured as one underlying
cause for long range correlations in the human brain. 

On the modeling front, recent models involving self-feedback loops in
simple motif networks have shown that coupling architecture may
improve long range correlations. Traub et al \cite{TraubWSJ96} uses a
linear self-feedback loop in the coupling architecture
of cortical neurons to inhance synchrony.  
In \cite{Klein06}, theory and experiments on semi-conductor
lasers, which exhibit bursting phenomena similar to neurons, show
that complete synchronization may occur with self-feedback loops. In
addition, lasers coupled in a line without self-feedback have been
shown to synchronize the end lasers \cite{FishcervbpMtG06,landsman06}, leading to
the possibility of long range correlations with delay.

Synchronization of nearby cells is often the result of receiving
common input, such as when a pyramidal cell sends projections to a targeted
area in the cortex.  While pyramidal cells tend to target specific areas,
matrix projection cells from the thalamus reach in a diffuse manner into
adjacent cortical areas helping to synchronize the activity of large
populations of cells.
Since many strong synaptic connections between different brain areas
tend to be one-way, 
many of the networks in the brain possess a hierarchical structure, which to
a first approximation is a forward network, modulated by a weaker modulatory
feedback (\cite{crick98}).  For example, there is a strong forward projection
from the thalamic LGN area to V1 or from V1 to MT, with weaker modulatory
feedback projections that modulate the magnitude of the cell's response 
(\cite{koch}).  
It has been suggested that exessively
strong mutually coupled loops in the brain 
would promote uncontrollable oscillations, such as in 
epilepsy (\cite{koch99}).  

With this in mind, we propose a
model of synchronization of cortical cells receiving a common time-delayed
input from a different area of the brain, such as the case with
projections of pyramidal cells to other cortical areas or the thalamus to the
cortex, and receiving a weaker time-delayed feedback.  
The central dynamical question addressed considers when the
synchronous behavior of such a neural network is stable,
particularly in regard to the synaptic coupling strengths, and the synaptic
time constant.  We find that shorter synaptic time constant of the target
cells promotes synchronization, and even increases the firing rate, 
for the same strength of input.  

\begin{figure}[ht]
\hspace*{-1 cm}
{
\epsfxsize=6in
\epsffile{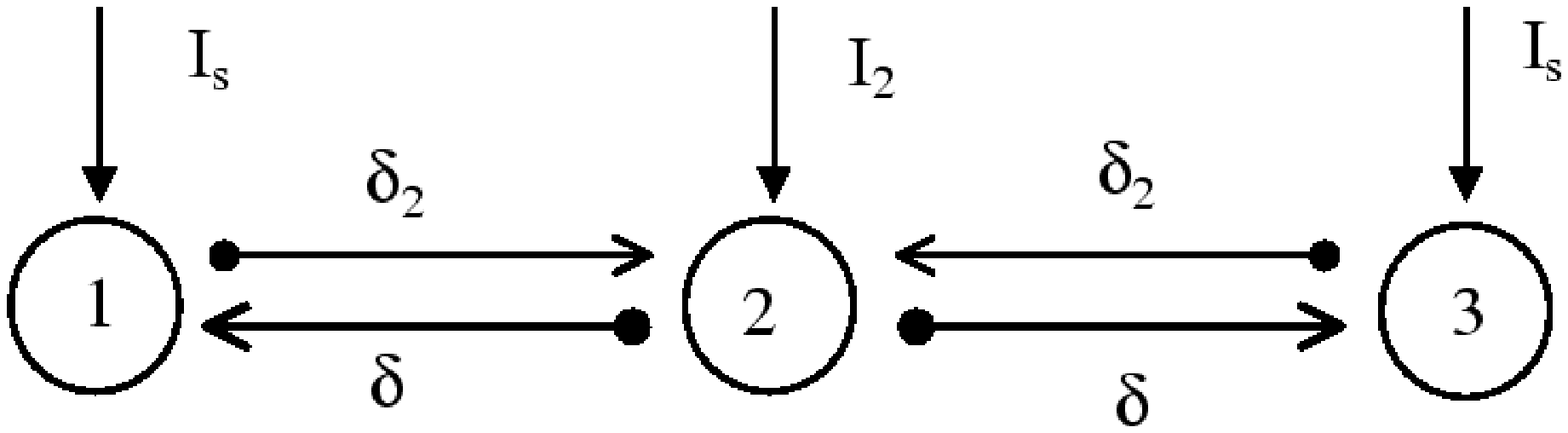}
}
\caption{The basic model of three synaptically coupled neurons.  The strength
of the coupling from the middle to the outer neurons is given by $\delta$
and from the outer to the middle by $\delta_2$.  The injected currents are
shown.  All synaptic coupling has propagation delay of $\tau_d$ and a synaptic
time constant of $\tau_{syn}$ and $\tau_{syn2}$ for the outer and the middle
neurons, respectively.}
\label{fig:setup}
\end{figure}

The paper is organized as follows: In Section II, the basic model is set up,
(see Fig. \ref{fig:setup})
and presented along with numerical results that show the dependence of
synchronization and firing rate on the strength of coupling and synaptic
conductance.  Section III, analyzes the synchronization observed for short
synaptic times by linearizing about the dynamics of two nearby trajectories.
It shows that fast synaptic input tends to synchronize two nearby
trajectories, regardless of their phase with respect to the input signal.
Section III concludes and summarizes.

\section{\label{sec:Model} II. Basic Model and Numerics}

\begin{figure}[ht]
\hspace*{-1 cm}
{
\epsfxsize=6in
\epsffile{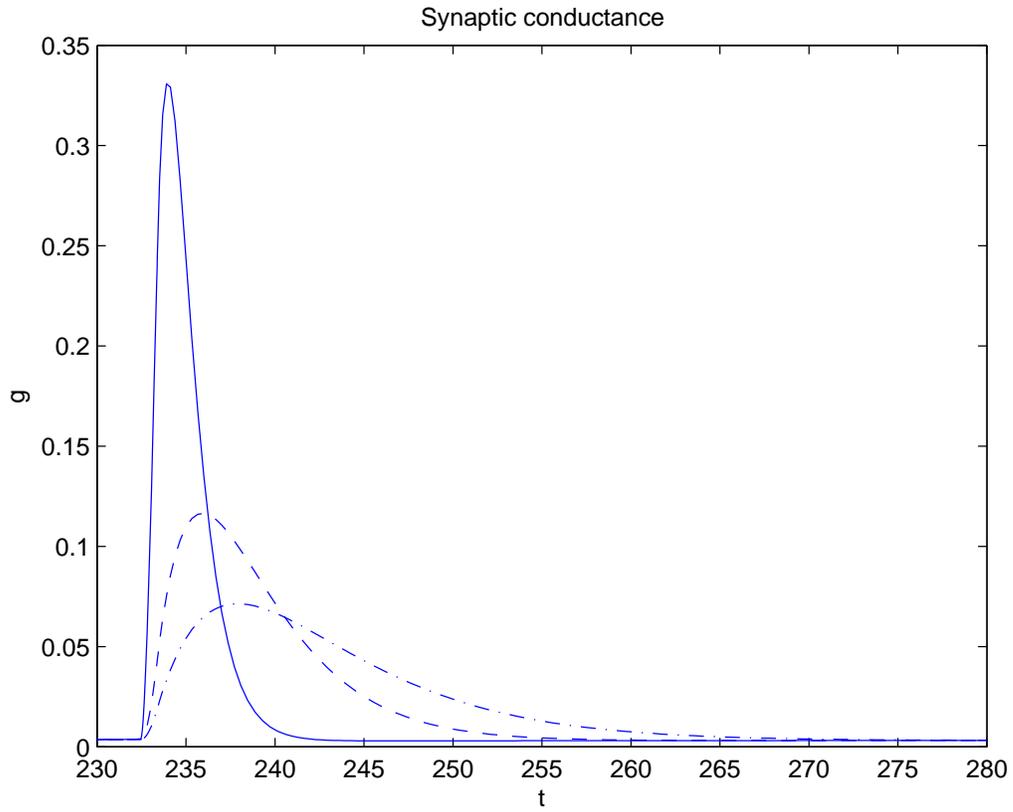}
}
\caption{Conductance, $g$, 
after a single pre-synaptic spike.  $\tau_{syn} = 1, 3,$
  and $5$ ms.}
\label{fig:g}
\end{figure}

Neurons can be largely divided into two classes, dependent on
their spiking properties.  Class I neurons can be stimulated to
fire at an arbitrarily low frequency, due to a saddle-node 
bifurcation, with increasing frequency as the magnitude of the
stimulating current increases. 
Class II neurons, on the other hand, only begin to fire at relatively
high frequency, with their limit cycle resulting from a subcritical
Hopf bifurcation.  Class II neurons are well represented by the
squid axon, while a large majority of the mammalian neurons are of
the Class I type.
Dynamics of a human neo-cortical neuron in the absence of
synaptic connections are well approximated by the following
Class I neuron equations: \cite{wilson}
\begin{displaymath}
\frac{dV}{dt} = -\{17.81 + 47.58 V + 33.8 V^2\}\left(V-0.48\right) - 26
R\left(V+0.95\right)+ I = F\left(V,R\right) + I 
\end{displaymath}
\begin{equation}
\frac{dR}{dt} = \frac{1}{\tau_R} \left(-R + 1.29 V + 0.79 +
3.3\left(V+0.38\right)^2\right) = G\left(V,R\right),
\label{eq:neuron}
\end{equation}
where $V$ and $R$ are voltage and recovery variables, respectively,
and $\tau_R=5.6$ ms.  The $dR/dt$ equation is written as a sum of a linear
term, for the normal $Na^+$ and $K^+$ currents, and a  quadratic terms in $V$
to approximate the the transient potassium current contributions \cite{rose89}.
The above model 
has been optimized to provide an accurate quantitative fit to the shape of a
regular spiking neuron potentials obtained from human neocortical neurons
\cite{foehring91}. 
Figure \ref{fig:limit} shows the limit cycle of a cortical neuron in
Eq. (\ref{eq:neuron}) 
when the applied  current, $I$, is above the bifurcation value,
resulting in a saddle-node bifurcation.
Due to a quadratic term in the recovery variable, the spike rate of a cortical
neuron can be arbitrarily low, for low currents, and increases as $I$ is
increased.  The above neurons can be coupled by adding an additional 
term to $dV/dt$ in Eq. (\ref{eq:neuron}) 
proportional to $g\left(V-E_{syn}\right)$, where $g$
is the synaptic conductance variable, $E_{syn}=0$ for excitatory synapses
and $-0.92$ for inhibitory.  The synaptic conductance, $g$, is obtained from
the following equations, commonly used for synaptic coupling \cite{wilson}, 
\begin{displaymath}
\frac{df}{dt} =  \frac{1}{\tau_{syn}} \left(-f + H_{step}\left(V_{pre} - \Omega\right)\right)
\end{displaymath}
\begin{equation}
\frac{dg}{dt} = \frac{1}{\tau_{syn}} \left(-g + f\right)
\label{eq:synaptic}
\end{equation}
where $H_{step}\left(x\right) = 1$, if $x>0$ and zero if $x<0$,
$V_{pre}$ is the voltage of the presynaptic neuron and
$\tau_{syn}$ is the synaptic conductance. In numerical simulation,
$\Omega = -0.20$ mV was chosen \cite{wilson}.
The reason for using two synaptic equations is that, depending on
$\tau_{syn}$, the conductance will peak after $V_{pre}$, continuing
to depolarize the membrane after the end of the presynaptic spike 
(see Figure \ref{fig:g}).  This type of response is consistent with
physiological data.  
For brief stimulus spike at $t=0$,
the two synaptic equations produce a response proportional to \cite{koch99}
\begin{equation}
g = \left(t/\tau_{syn}^2\right) exp\left(-t/\tau_{syn}\right)
\label{eq:g}
\end{equation}
Figure \ref{fig:g} shows $g$ for different 
values of $\tau_{syn}$.  In each case, the area under the curve is the
same (equal to one), with $\tau_{syn}$ determining the width of
the post-synaptic spike.

Using Eqs. (\ref{eq:neuron}) and (\ref{eq:synaptic}) we can now set up
the basic model of three mutually coupled neurons.
The basic set-up is shown in Figure \ref{fig:setup}, where 3 neurons
are coupled in a line.  The middle neuron sends an excitatory time-delayed
signal to the two outer neurons which are coupled to the middle one via 
weaker delayed coupling with a longer synaptic time constant.  This model 
reflects the
typically observed hierarchy in the brain (described in the introduction), 
where the neurons in one brain area,
such as the thalamus, send strong excitatory connections to cortical neurons,
receiving weak modulatory feedback.  Since the equations of the outer neurons
are identical, receiving the same synaptic input from the middle cell, their
synchronization would indicate that a group of identical neurons fire
synchronously when subjected to a particular synaptic input.  
Using Eq. (\ref{eq:neuron}), the 
equations for the coupled neurons are given by:
\begin{displaymath}
\frac{dV_i}{dt} = F\left(V_i, R_i \right) + I_i 
- \delta_i g_i \left(V_i - E_{syn}\right)
\end{displaymath}
\begin{equation}
\frac{dR_i}{dt} = G\left(V_i,R_i\right)
\label{eq:coupled}
\end{equation}
where $i$ is the index of each neuron, with outer neurons receiving
the same current, $I_{1,3} \equiv I_s$, and the
same synaptic strength, $\delta_1 = \delta_3 \equiv \delta$.  
The current input to the middle neuron is higher than to the outer ones: $I_2
> I_s$, leading to the lower uncoupled spiking rate for the outer cells.  This
was done so that the higher activity of the inner cell drives the outer ones,
as might be the case in a typical hierchical network.
The synaptic input to the middle neuron is weaker, $\delta_2 < \delta$.  
The synaptic time-constant in 
Eq. (\ref{eq:synaptic}) is the same for the outer neurons,
$\tau_{syn1}=\tau_{syn3} \equiv \tau_{syn}$,
but longer for the middle neuron, to model the affect of a slowly-varying
modulatory feedback.  
The conductances, $g_i$, are obtained from
Eq. (\ref{eq:synaptic}), with presynaptic voltage to the middle cell
given by:
$V_{pre2} = V_1\left(t-\tau_d\right) + V_3\left(t-\tau_d\right)$, 
and to the outer cells by $V_{pre1,3} = V_2\left(t-\tau_d\right)$.  The delay
in the propagation of the signal is given by $\tau_d$. 

\subsection{\label{sec:subsec1} The effect of delays on correlation and
  synchronization }
\begin{figure}[ht]
\hspace*{-1 cm}
{
\epsfxsize=6in
\epsffile{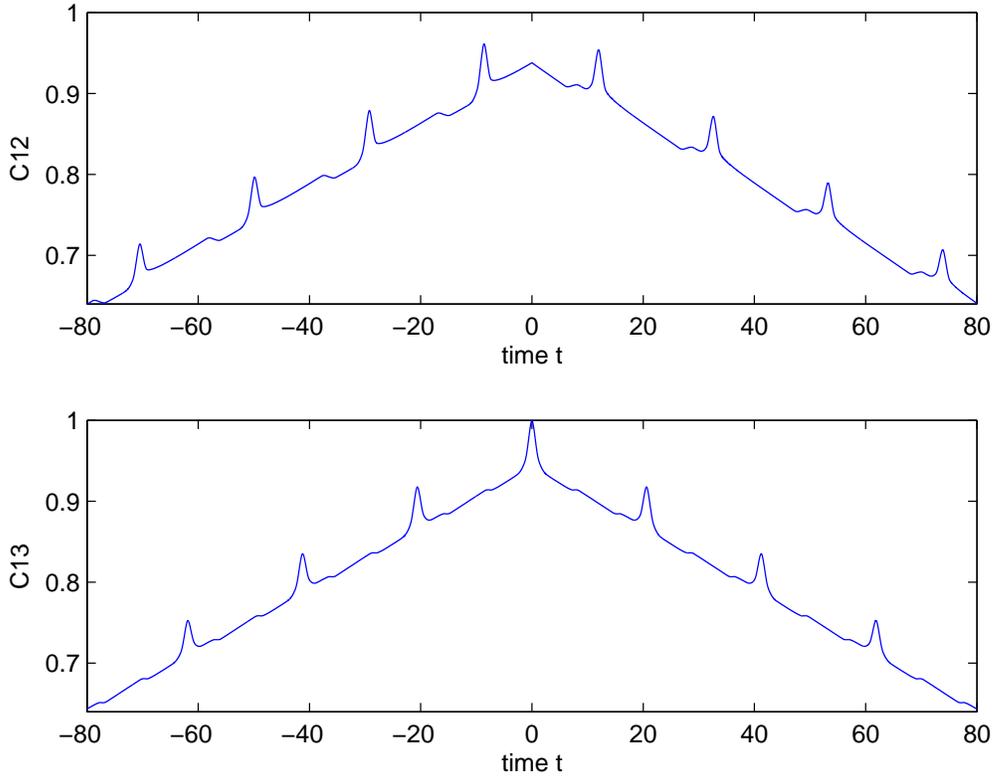}
}
\caption{Correlations of spiking output.  Top: between the outer and the
  middle neuron.  Bottom: Between outer neurons.  $\tau_d=10$ ms, $\delta=4$,
  $\delta_2 = 2$, $\tau_{syn}=1$ ms, $\tau_{syn2}=2$ ms, $I_{1,3}=I_s=0.22$ nA,
  $I_2=0.5$ nA.}
\label{fig:correlations}
\end{figure}

While the length of the synaptic delay, $\tau_d$, does not seem to affect the
degree of synchronization, it has a substantial affect on correlations and
phase relations between neurons.  Figure \ref{fig:correlations} shows
correlations in spiking output between the middle and outer neuron, and
between outer neurons for $\tau_d=10$.  
 The $x$-axis indicates the time-shift at which the
correlations function was computed.
The outer neurons are synchronized,
since $C_{13}=1$ at $t=0$.   It can be seen that the greatest
correlations between the inner and outer neurons occur when $t$ is shifted by
the delay time, $\tau_d$.  The delay also creates spikes in correlation at
intervals of twice the delay time, as can be seen in
Fig. \ref{fig:correlations}.  Since, in this case, the outer neurons are
synchronized, the spikes indicate that there are self-correlations 
in the time series of a single neuron at intervals of $2 \tau_d$.  This is the
round-trip, or feed-back, time, since its the minimum
time that it would take for a signal to travel
from one of the neurons, affect the target, and get back to that same neuron.  
It follows that time delays lead to self-correlations in the spike trains
that are not observed when delays are absent, and which may lead to
more regular patterns in the time-series data.  

For reasons explained in the introduction, 
$\delta > \delta_2$ and $\tau_{syn} < \tau_{syn2}$  
were used to model the affect of
a stronger 
forward and a weaker modulatory feedback.  For this type of coupling,
the middle neuron leads the outer ones by the delay time, $\tau_d$.  This
type of phase-locking behavior has been observed in other time-delay systems,
such as lasers \cite{HeilFEMM01,WhiteMM02}.  Thus the time-shift between two correlated
spike trains should be directly
related to the delays in transmission, with the input
leading the output by the delay time.

\subsection{\label{sec:subsec2} The effect of the coupling strength on firing rate and synchronization}
\begin{figure}[ht]
\hspace*{-1 cm}
{
\epsfxsize=6in
\epsffile{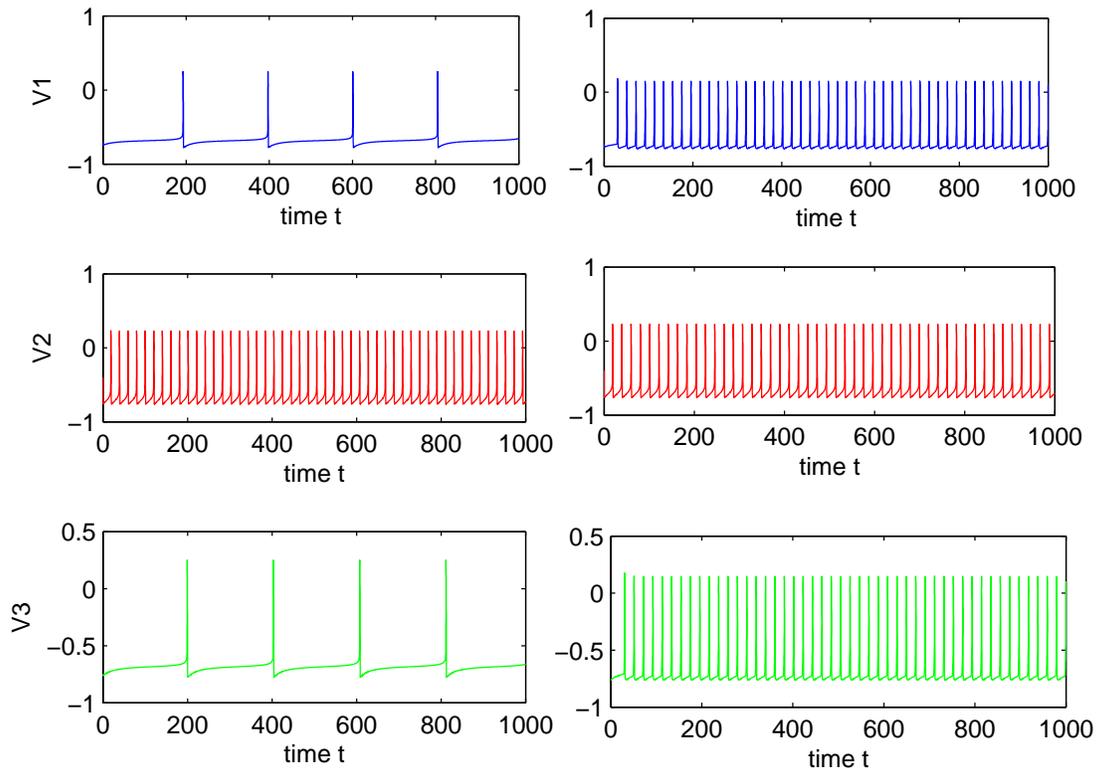}
}
\caption{Left: spiking in the absence of coupling.  Right: coupling,
  $\delta=4$, $\delta_2 = 2$, $\tau_{syn}=1$,  $\tau_{syn2}=2$.  
In both cases, $I_s = 0.22$, $I_2 = 0.5$.}
\label{fig:uncoupled}
\end{figure}

\begin{figure}[ht]
\hspace*{-1 cm}
{
\epsfxsize=6in
\epsffile{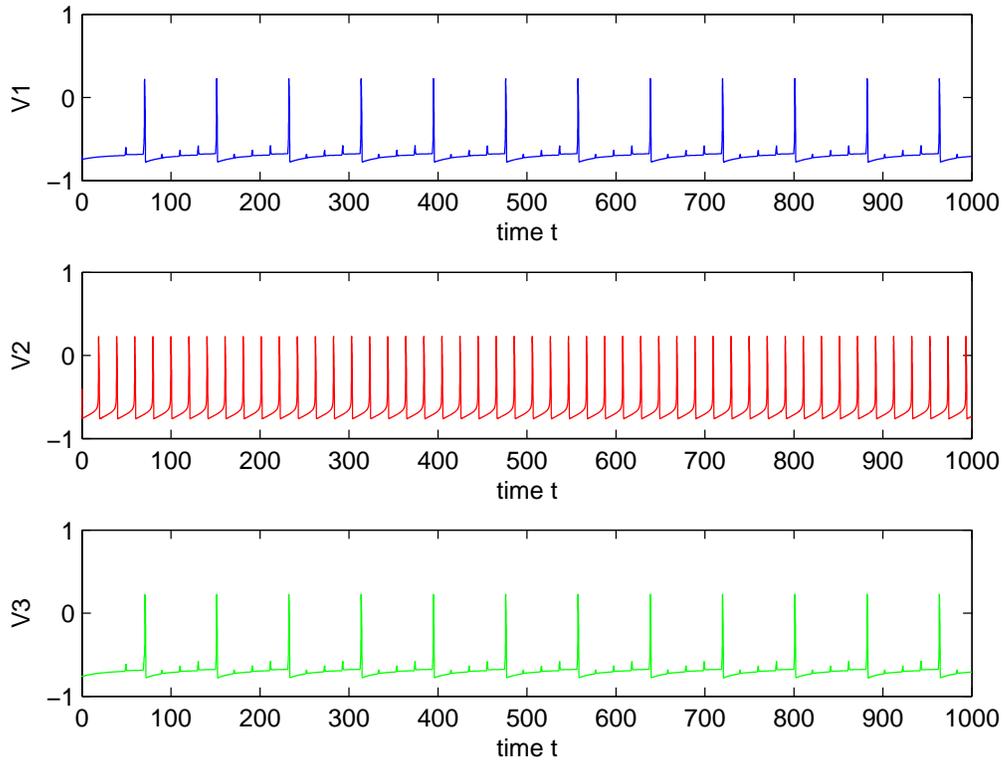}
}
\caption{Spiking voltage for weak coupling, 
$\delta=0.5$, $\tau_{syn}=0.03$, $\delta_2=0.3$. 
There is $1:4$ frequency locking between the outer and the middle neurons.}
\label{fig:firing_rate}
\end{figure}

\begin{figure}[ht]
\hspace*{-1 cm}
{
\epsfxsize=6in
\epsffile{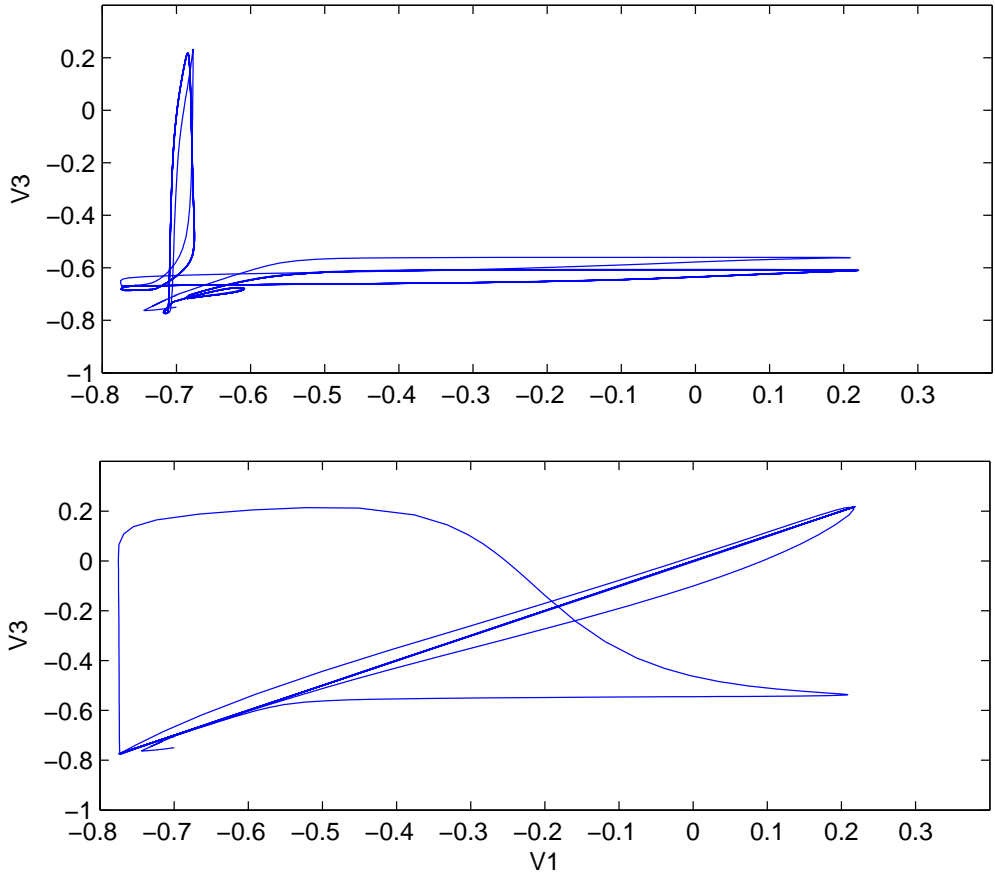}
}
\caption{Sensitive dependence of synchronization on coupling strength.
Top: $\delta=1.02$, Bottom: $\delta=1.03$, $\tau_{syn}=0.5$.}
\label{fig:delta}
\end{figure}

Increasing the synaptic coupling in general increases the firing rate of the
neurons for excitatory synapses.  Above a certain value of the coupling
strength, there is phase-locking between the inner and the outer neurons,
whereby all firing rates are equal.  The bifurcation value of synaptic
coupling at which $1:1$ frequency locking occurs depends on the current
injected into each cell.  In the absence of coupling, a significant 
difference in injected current for the outer and inner
neurons,  ($I_s = 0.22$ and $I_2 = 0.5$ in simulations) 
leads to a big difference in firing rate. Figure \ref{fig:uncoupled} shows a typical voltage trace for the uncoupled, 
$\delta=0$,
and coupled neurons, $\delta=4$, where coupling is sufficiently strong to cause
phase-locking.    Phase-locking or $1:1$ frequency locking occurs for
$\delta > 3.4$.  
This type of behavior has been observed for mutually coupled
neurons in the absence of delays \cite{wilson}.  For lesser values of the
coupling strength, different frequency locked behaviors are observed, with
frequency ratio between $1$ and $f_i/f_o$, where $f_i$ and $f_o$ are the
frequencies of inner and outer neurons, respectively, in the absence of
synaptic coupling.  Figure \ref{fig:firing_rate} shows $1:4$ frequency locking
that occurs for weak synaptic coupling: $\delta = 0.5$, $\delta_2 = 0.3$.
When the coupling is too weak, the outer neurons become desynchronized, as
shown in Figure \ref{fig:delta}. 
The bifurcation value of $\delta$, however
depends on the synaptic time constant, $\tau_{syn}$, of the outer-neurons.  
Thus Figs. \ref{fig:firing_rate} and  \ref{fig:synch_limit}, 
which have a very short time constant of $\tau_{syn}=0.03$, 
shows synchronization at a lower coupling strength of $\delta=0.5$, 
compared to a minimum $\delta=1.03$ needed for synchronization of 
neurons in Fig. \ref{fig:delta}, 
where $\tau_{syn}=0.5$ (a more realistic value for fast synapses).  
This sensitive dependence of synchronization on the coupling strength and 
synaptic time constant of the 
outer-neurons is explored analytically in Section III.

\subsection{\label{sec:subsec3} The effect of synaptic time-constant on
  synchronization and firing rate}
\begin{figure}[ht]
\hspace*{-1 cm}
{
\epsfxsize=6in
\epsffile{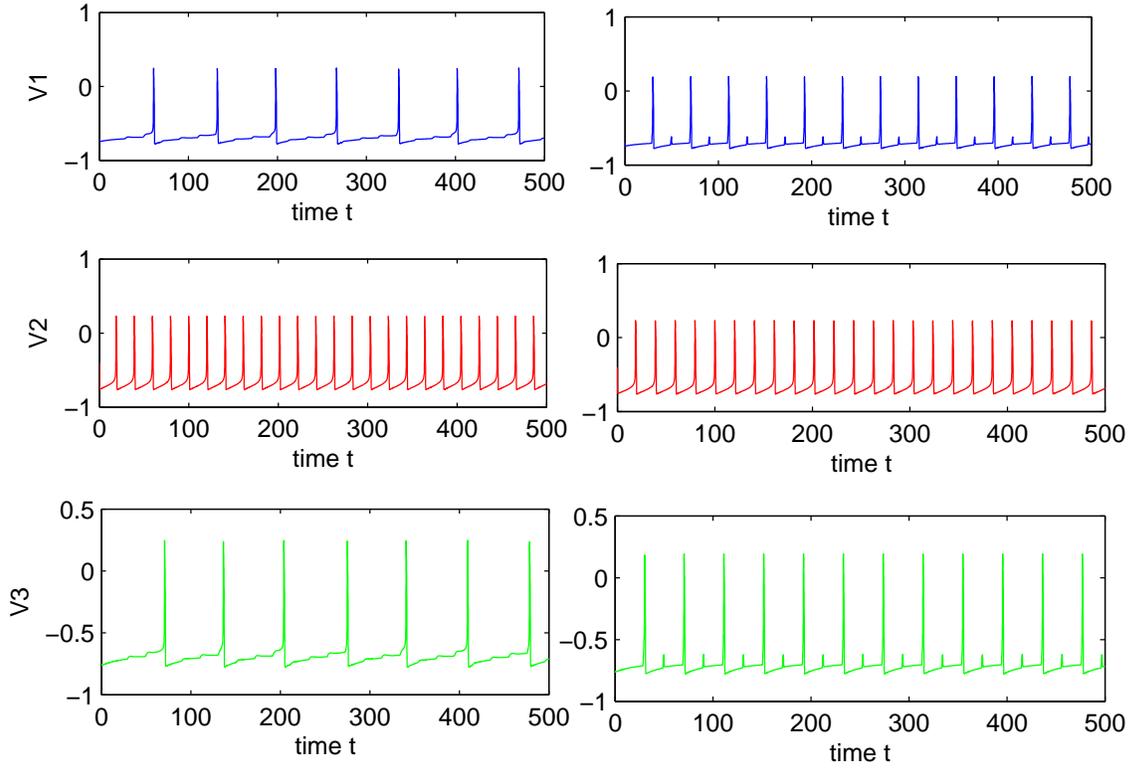}
}
\caption{Increase in firing rate as $\tau_{syn}$ decreases.
Left: $\tau_{syn} = 2$.  Right: $\tau_{syn} = 0.1$.  
In both cases, $\delta=1$,   $\delta_2=0.2$, $\tau_{syn2} = 4$, $I_s = 0.22$, $I_2 = 0.5$.}
\label{fig:gfiring_rate}
\end{figure}

\begin{figure}[ht]
\hspace*{-1 cm}
{
\epsfxsize=6in
\epsffile{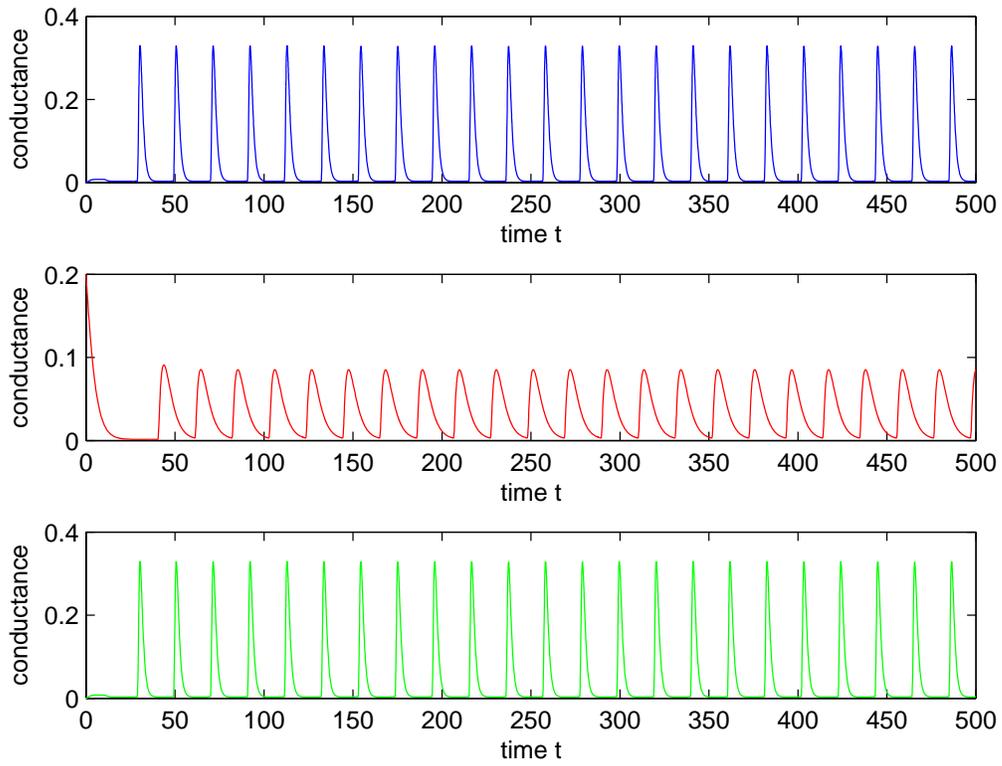}
}
\caption{Conductance for $\tau_{syn}=1$, $\tau_{syn2}=3$.  All else as in Fig. \ref{fig:correlations}}
\label{fig:full_g}
\end{figure}

Figure \ref{fig:gfiring_rate} shows an increase in firing rate of the outer
neurons as their synaptic time constant, $\tau_{syn}$, is decreased.
This increase in firing rate may be surprising, 
since the time constant only controls the 
width of the conductance spike and not the area under the curve,
as shown in Fig. \ref{fig:g}.  Thus the contribution of a pre-synaptic
spike to a change in post-synaptic voltage during an inter-spike interval
is largely independent of the synaptic time constant. This can be seen
by integrating the $g\left(V_i-E_{syn}\right)$ term in Eq. (\ref{eq:coupled})
over the interval of conductance change.  
Figure \ref{fig:full_g} shows a fluctuation in  conductance for
a system given by Eqs. (\ref{eq:synaptic}) and (\ref{eq:coupled}), with
$\tau_{syn} = 1$ and $\tau_{syn2} = 3$.
A lower firing rate that occurs for slower synapses may be the
result of the decay of any increase in voltage during the inter-spike
interval back to the limit cycle trajectory (see the bottom of
Fig. \ref{fig:synch_limit}).  A pre-synaptic spike from the inner neuron can
trigger a spike from the outer one, when it is delivered toward the end of the
inter-spike interval.  Thus a more narrow jump in conductance and the
resultant jump in voltage, $V_{1,3}$, may mean that there is less decay before
the critical threshold is reached, thereby increasing the likelihood of a
spike. 

\begin{figure}[ht]
\hspace*{-1 cm}
{
\epsfxsize=6in
\epsffile{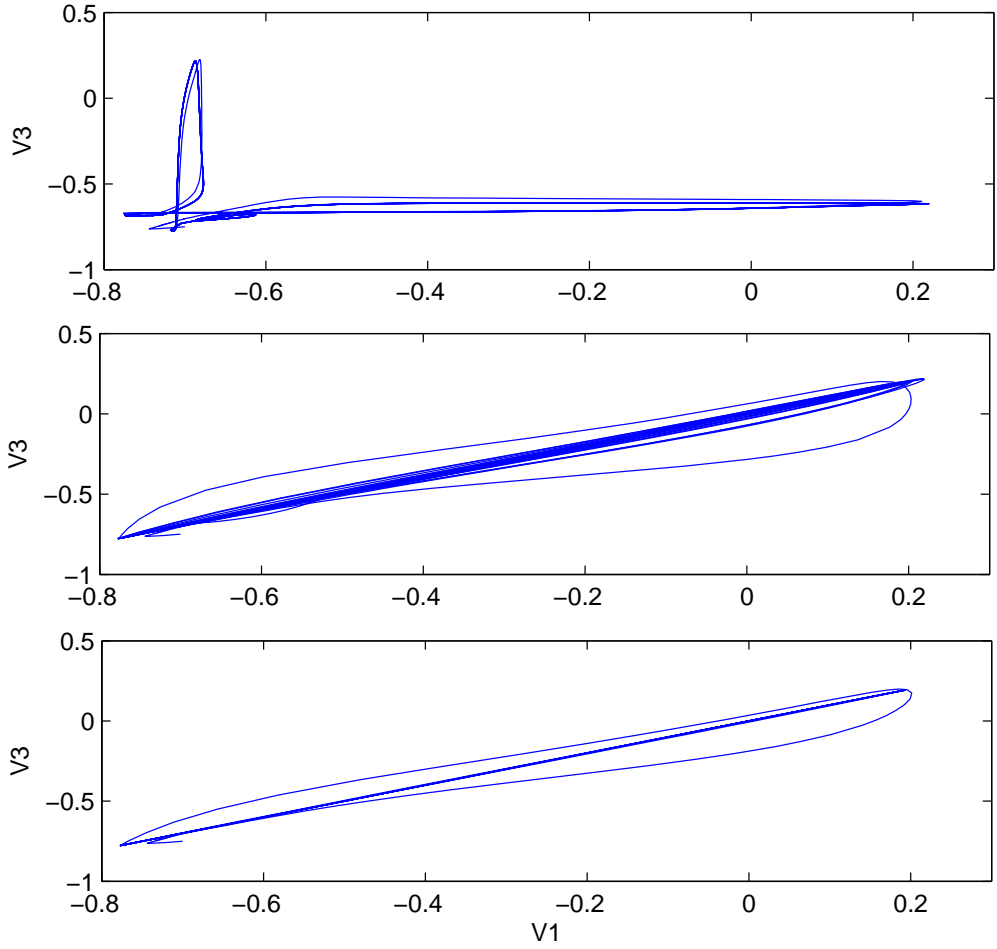}
}
\caption{Dependence of synchronization on the synaptic time constant,
  $\tau_{syn} $.
Top: $\tau_{syn} = 0.5$, Middle: $\tau_{syn} = 0.3$, 
Bottom: $\tau_{syn} = 0.2$, In all three
cases, $\delta = 1$, $\delta_2 = 0$.}
\label{fig:compare_tau}
\end{figure}

The most noticeable affect of the decrease in $\tau$ is greater
synchronization.  This is shown in Fig. \ref{fig:compare_tau}, where the outer
neurons become progressively synchronized as $\tau_{syn}$ is 
decreased from $0.5$ to $0.2$.  It can be seen that for
relatively weak coupling of $\delta = 1$,
$\delta_2 = 0$, and  $\tau_{syn} = 0.2$, the outer neurons are completely
 synchronized, after the transients die out.  The next section analyzes the
 affect of a presynaptic spike on a fast synapse (small $\tau_{syn}$)
 in synchronizing two nearby trajectories of the outer neurons.

\section{\label{sec:Analysis} III. Analysis of synchronization for weakly coupled,
fast synapses}
\begin{figure}[ht]
\hspace*{-1 cm}
{
\epsfxsize=6in
\epsffile{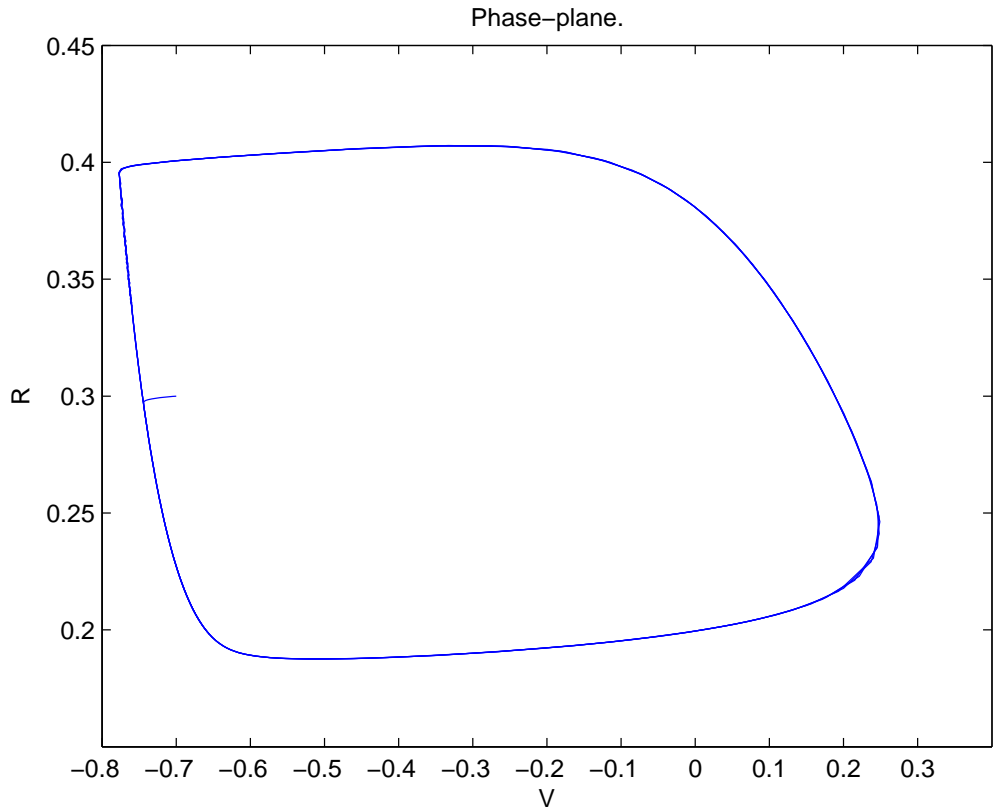}
}
\caption{Limit cycle of an uncoupled cortical neuron, given by
  Eq. (\ref{eq:neuron}). $I=0.5$ nA.}
\label{fig:limit}
\end{figure}

The three neuron model described by Eqs. (\ref{eq:synaptic}) and
(\ref{eq:coupled}) possesses internal symmetry, e.g. its equations of
motion do not change if the variables $\{V_1,R_1\}$ and  $\{V_3,R_3\}$ are
interchanged.  It follows that the synchronized
regime, where the symmetric variables are
exactly equal: $\{V_1=V_3=V_0, R_1=R_3=R_0\}$, 
is a solution \cite{pikovsky}.  
For an uncoupled
system, this solution would not be a stable one, since any perturbation along
the limit cycle would result in a phase-difference.  A spiking input, $V_2$,
from the middle neuron stabilized the synchronous state 
when the synaptic time-constant, $\tau_{syn}$ is sufficiently short.  
To study the affect
of a single pre-synaptic spike on two identical neurons, with nearby
trajectories along the limit cycle, introduce new variables:
$\tilde V = V_1 - V_3$ and  $\tilde R = R_1 -R_3$.  These new variables
correspond to a perturbation transverse to the synchronized state: 
$\{V_0, R_0\}$.
Using Eqs.  (\ref{eq:neuron}), (\ref{eq:g}) and (\ref{eq:coupled}), the
linearized dynamics are:
\begin{equation}
\frac{d \tilde V}{dt} = - n\left(V_0\right) \tilde R - h\left(V_0,R_0\right)
\tilde V  - \delta \left(t_s/\tau_{syn}^2\right) exp\left(-t_s/\tau_{syn}\right) \tilde V
\label{eq:diffV}
\end{equation}

\begin{equation}
\frac{d \tilde R}{dt} = \frac{1}{5.6} \left(- \tilde R - m\left(V_0,R_0\right)
\tilde V\right)
\label{eq:diffR}
\end{equation}
where $n\left(V_0 \right)=26 \left(V_0 + 0.95\right)$,
$h\left(V_0,R_0\right)=\left(-5.03 + 101.4 V_0^2 - 32.45 V_0 + 26
R_0\right)$, and $m\left(V_0,R_0\right)=-\left(3.8 + 6.6 V_0 \right)$.
$t_s$ denotes the time since the last spike of the center neuron.
Equation (\ref{eq:diffV}) is valid for a short synaptic time constant,
when only the effect of the last presynaptic spike is
significant (see Fig. \ref{fig:g}).
During an inter-spike interval of the 
outer cells, $n, h$ and $m$ are all positive. 
This can be easily confirmed by using
Fig. \ref{fig:limit}, where $-0.8 < V < -0.6$ and $.18 < R < 0.4$ during the
inter-spike interval, and calculating the lowest possible values of $n,h$ and
$m$ for a given range of $V_0$ and $R_0$.
Since $\{\tilde V, \tilde R\}$ denote the difference between the two nearby
trajectories along the limit cycle,  there is a
relationship between the two variables given by 
\begin{equation}
\tilde R(t) = - l(V_0)  \tilde V(t)
\label{eq:slope}
\end{equation}
where $l$ is the negative of the
slope of the limit cycle at $\{V_0(t), R_0(t)\}$.  From
Fig. \ref{fig:limit}, $l(V_0) \approx 2$ during the inter-spike interval.  

A change in $\tilde V$ following a post-synaptic spike can now be calculated 
by substituting Eq. (\ref{eq:slope}) into Eq.  (\ref{eq:diffV}), dividing 
by $\tilde V$ and integrating over $\triangle t$, 
the duration of the conductance spike, $\left(t_s/\tau_{syn}^2\right)
exp\left(-t_s/\tau_{syn}\right)$.  
For a short synaptic time constant, $\tau_{syn} \leq 1$, 
the width of $\triangle t$ is small compared
to the time-scale of neuronal dynamics during the inter-spike interval of the
outer cells.  A change in $\tilde V$ following a post-synaptic spike is now
given by, 
\begin{equation}
\tilde V\left(t+\triangle t\right) = 
\tilde V_u\left(t+\triangle t\right) \cdot e^{-\delta} 
\label{eq:tildeV}
\end{equation}
where $\tilde V_u\left(t+\triangle t\right)$ is the voltage difference of
two nearby trajectories at time $t+\triangle t$ in the absence of synaptic
input:
 $\tilde V_u\left(t+\triangle t\right) \approx \tilde V\left(t\right)
exp\left(\triangle t\left(n l - h\right)\right)$, with $n$ and $h$ defined
after Eq. (\ref{eq:diffR}) and $l$ given in Eq. (\ref{eq:slope}).
For small $\triangle t$, so that $|\triangle t\left(n l - h\right)| 
\ll \{1,\delta\}$,
most of the change in voltage during a post-synaptic spike is caused by the
spike itself so that Eq. (\ref{eq:tildeV}) can be approximated as
\begin{equation}
\tilde V\left(t+\triangle t\right) 
\approx  \tilde V\left(t\right) e^{-\delta} 
\label{eq:tildeV2}
\end{equation}
Using Eqs. (\ref{eq:diffR})-(\ref{eq:tildeV}), the dynamics of $\tilde R$
immediately following a narrow post-synaptic spike can be approximated as
\begin{equation}
\frac{d \tilde R}{dt} = \frac{1}{5.6}\left(- 1 +
\left(\frac{m\left(V_0,R_0\right)}{l\left(V_0\right)}\right) e^{-\delta}\right) \tilde R
\label{eq:tildeR}
\end{equation}
From Eqs. (\ref{eq:tildeV}) and (\ref{eq:tildeV2}), 
a single presynaptic spike
from a center neuron acts to
decrease the voltage difference, $\tilde V$, of the outer neurons by a
factor of $exp\left(-\delta\right)$.  This presynaptic spike also
decreases $\tilde R$ by
decreasing the positive contribution from the $m/l > 0$ term in
Eq. (\ref{eq:tildeR}).  
Thus the effect of the presynaptic input is to
decrease the perturbation from synchronized state of the outer neurons,
pushing their trajectories closer in phase-space.  

Eqs. (\ref{eq:tildeV})-(\ref{eq:tildeR}) show that a spiking input from
the middle neuron has a stabilizing affect on the synchronized state when
the synaptic time constant is short.  Since the difference in trajectories,
$\tilde V, \tilde R$ is taken along a limit cycle, the maximum Lyapunov exponent in
the absence of synaptic coupling would be zero (corresponding to the
displacement along a trajectory in phase-space), and the transverse
exponents must be negative since the systems is dissipative.  It follows that, for
sufficiently small $\tau_{syn}$,  
a common synaptic input acting during the inter-spike interval should
eventually synchronize the neurons, even for weak synaptic coupling.

\begin{figure}[ht]
\hspace*{-1 cm}
{
\epsfxsize=6in
\epsffile{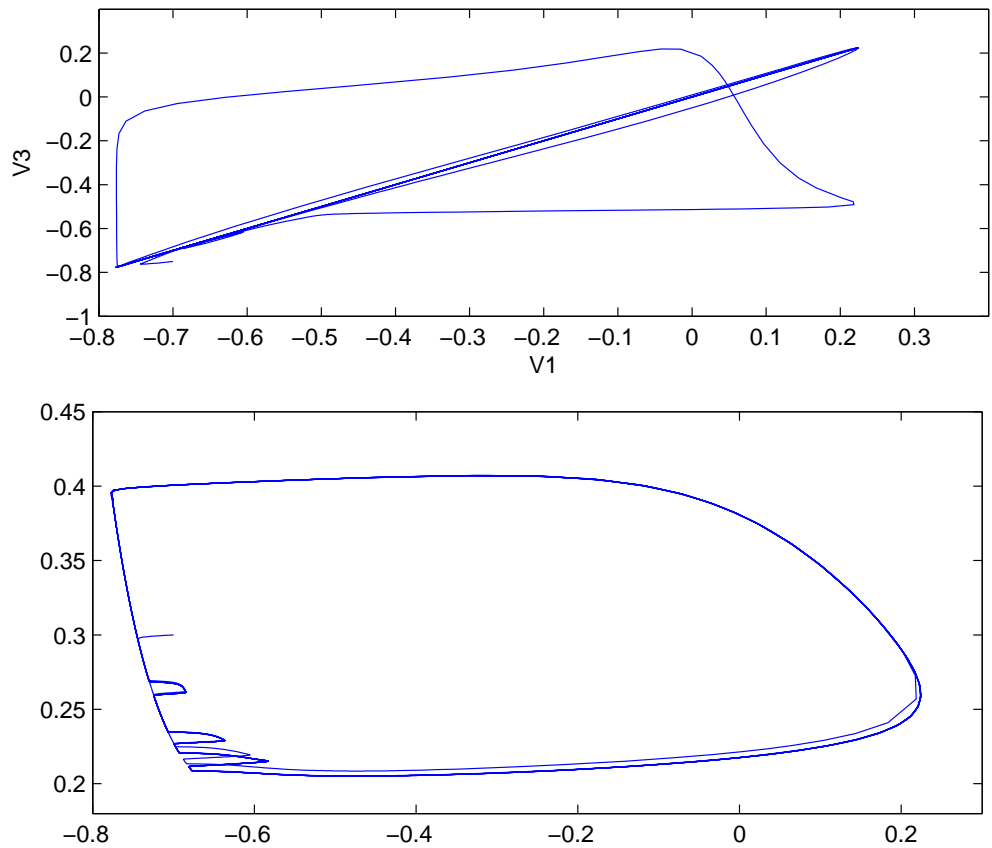}
}
\caption{Synchronization for weakly coupled, fast synapses.
$\delta=0.5$, $\tau_{syn}=0.03$, $\delta_2=0.3$.  
Top: synchronization
after the transients die out.  Bottom: Limit cycle of one of the outer
neurons.  Same parameters as in Fig. \ref{fig:firing_rate}.
}
\label{fig:synch_limit}
\end{figure}

Figure \ref{fig:synch_limit} shows the synchronization of the outer neurons
for a very short synaptic time constant, $\tau_{syn}=0.03$ and weak coupling,
$\delta = 0.5$.  After the transients die out, the outer neurons become
synchronized, thereby falling on a straight line in the $V_1$ vs $V_3$ plot.
 As can be seen in Figure \ref{fig:firing_rate}, 
for weak coupling, there is a big difference in firing
rate between the outer and the middle neuron, 
due to differences in injected current.
It follows that the synchronization is not due to phase-locking between the
middle and the outer cells.  The effect of the spiking input on the limit
cycle trajectory can be seen at the bottom of Figure  \ref{fig:synch_limit}).  
There is an 
integer ratio between the inner and the outer frequencies, whereby the
onset of a spike in the outer neuron is triggered by spiking input from the
middle one, delivered toward the end of the inter-spike interval.  The
phenomena is similar to the subharmonic resonance where the limit cycle
responds at a subharmonic of the stimulus frequency \cite{wilson}.

Eqs. (\ref{eq:tildeV})-(\ref{eq:tildeR}) show a strong dependence
of synchronization on the coupling strength.  This is confirmed by numerics.
Figure \ref{fig:delta} shows $V_1$ vs $V_3$ for two slightly different values
of $\delta$, $\delta=1.02$ and $\delta=1.03$.  A slight change in $\delta$
leads to an onset of synchronization between the outer neurons.    

\section{\label{sec:concl} IV. Conclusion}
A model of three mutually coupled cortical neurons with delays was studied 
using analysis and numerical simulation.  The outer neurons were stimulated
with smaller current and had a much lower frequency in the uncoupled case,
with their frequency significantly increasing depending on the strength of
synaptic coupling with the middle neuron.
At higher values of the synaptic
coupling constant, typical phase-locked behavior and $1:1$ frequency locking
was found between the middle and the outer neurons, with different frequency
locking ratios as the synaptic strength was lowered.  It was found that delays
affected the time-series data  by introducing correlations at the time-scale
of the delay.  While the spiking behavior in the synchronized case 
was fairly regular,
this effect would be interesting to explore for a more complicated, chaotic
spike train that can be achieved by incorporating slow adaptation currents
into the neuron model.
In the case of phase or frequency locking, the middle neuron leads 
the outer by the delay time, $\tau_d$.  

While synchronization of  outer neurons was sensitive to the
synaptic strength, the synaptic time constant for outer neurons, $\tau_{syn}$
was also highly significant.  It was found that shorter synaptic constant
substantially improves synchronization, able to synchronize the neurons even
when the coupling strength is very weak.  A short synaptic constant was also
able to induce $1:1$ frequency locking at a much weaker mutual coupling than
this type of dynamics might be expected, for the input currents used.   
Analysis of dynamics for fast synapses showed that spiking synaptic input
during the inter-spike interval stabilized the synchronization manifold,
even for arbitrarily weak coupling, and independent of the
phase relationship between the inner and outer cells.  This indicates that
even a very weak synaptic input can synchronize  cells, as long
as the synaptic time constant is sufficiently short.  The finding may have
significance in synchronizing large groups of cells in the cortex via weak
synaptic input from other areas, such as the thalamus, or other areas in the
cortex proper.


\section{Acknowledgments}
This work was supported by a grant from the Office of Naval
Research.  ASL
is currently a post doctoral fellow with the National
Research Council.

\pagebreak

\end{document}